\documentclass[pre,aps,twocolumn]{revtex4}
\usepackage{amsmath, amsfonts, amssymb}
\usepackage{graphicx,epsfig}
\usepackage{color}
\graphicspath{{images/}}

\newcommand{\del}{\partial}

\begin{document}

\title{Glassy states and superrelaxation in populations of coupled phase oscillators}

\author{D.\ Iatsenko, P.\ V.\ E.\ McClintock and A.\ Stefanovska}

\affiliation{Department of Physics, Lancaster University, Lancaster LA1 4YB,
United Kingdom}

\begin{abstract}
Large networks of coupled oscillators appear in many branches of science, so that the kinds of phenomena they exhibit are not only of intrinsic interest but also of very wide importance. In 1975, Kuramoto proposed an analytically tractable model to describe such systems, which has since been successfully applied in many contexts and remains a subject of intensive research. Some related problems, however, remain unclarified for decades, such as the properties of the oscillator glass state discovered by Daido in 1992. Here we present a detailed analysis of a very general form of the Kuramoto model. In particular, we find the conditions when it can exhibit glassy behavior, which represents a kind of synchronous disorder in the present case. Furthermore, we discover a new and intriguing phenomenon that we refer to as {\it superrelaxation} where, for a class of parameter distributions, the oscillators feel no interaction at all during relaxation to incoherence, a phenomenon reminiscent of superfluidity or superconductivity. Our findings offer the possibility of creating glassy states and observing superrelaxation in real systems, thus paving the way to a cascade of applications and further research in the field.
\end{abstract}

\maketitle

\section*{Introduction}

The Kuramoto model (KM) \cite{Kuramoto:84} was introduced and developed to provide an analytically tractable description of the populations of coupled phase oscillators that so often appear in real life. It has been applied successfully in many fields \cite{Strogatz:00,Acebron:05}, e.g.\ to describe the collective behavior of lasers \cite{Javaloyes:08,Oliva:01}, neurons in the brain \cite{Sheeba:08a}, Josephson junction arrays \cite{Wiesenfeld:98}, and even humans \cite{Neda:00}. The widespread practical applications of the KM has ensured that the basic model, together with its variants and modifications \cite{Iatsenko:13,Petkoski:12,Montbrio:06,Montbrio:11a,Pazo:11,Montbrio:11b,Skardal:12,Anderson:12,Lee:11,Hong:11a,Hong:11b,Hong:12,Skardal:11,Lee:09,Sakaguchi:86}, has been studied very extensively over the past decades. The model has been found to exhibit many interesting phenomena and states and, by now, most of them have been thoroughly investigated. There is, however, one important exception, namely the so-called {\it oscillator glass} state \cite{Daido:92,Bonilla:93,Stiller:98,Daido:00,Stiller:00} whose provenance and properties remain largely unresolved. First reported by Daido \cite{Daido:92} in 1992, this state could not be studied analytically due to the complexity of the model that was being used and, since then, it has not to our knowledge been found in any of the other KM variants.

Here we report the analysis of a very general, yet analytically tractable, form of KM. We derive a set of equations describing its steady state behavior, and show that, for a particular class of distributions, the consideration may be reduced to a much simpler model. We also discuss when and how the full time-evolution of parameters can be obtained, and find some interesting features related to this question. Among the phenomena that the model can exhibit we find glassy behavior, which can be studied analytically within the framework presented, and we discuss the glassy states and their properties in detail. Finally, we describe a completely new phenomenon which we refer to as \emph{superrelaxation} where, under certain conditions, the coupling between the oscillators effectively disappears during their relaxation to incoherence.

\section*{The model}

We consider the KM in the form
\begin{equation}\label{km}
\dot{\theta_i}=\omega_i-\frac{k_i}{N}\sum_{j=1}^Nq_j\sin(\theta_i-\theta_j-\beta_i-\gamma_j),
\end{equation}
where $N$ is the number of oscillators, $\theta_i(t)$ and $\omega_i$ are respectively the $i$th oscillator's phase and natural frequency, $k_iq_j$ ($\beta_i+\gamma_j$) represent the coupling strengths (phase lags) between the $i$th and $j$th oscillators. All parameters $\Gamma_i\equiv\{\omega_i,k_i,q_i,\beta_i,\gamma_i\}$ are drawn from a joint probability density $G(\Gamma)\equiv G(\omega,k,q,\beta,\gamma)$. The case considered earlier \cite{Iatsenko:13} corresponds to (\ref{km}) with $q_i=1,\beta_i=\gamma_i=0$.

The KM has not previously been treated in such a general form (\ref{km}), though it actually includes as special cases the many KM modifications and extensions studied earlier. Thus, not much is known about the possible behavior of the system except in those particular cases. To study it, we first generalize the recently presented framework \cite{Iatsenko:13} to encompass (\ref{km}), and then we proceed to a consideration of the phenomena it can exhibit.

\section*{Main equations}

The oscillators' collective behavior in (\ref{km}) can be described by two complex parameters
\begin{equation}\label{ZYdef}
Z\equiv Re^{i\Psi}\equiv\frac{1}{N}\sum_{j=1}^Ne^{i\theta_j},\;Y\equiv We^{i\Phi}\equiv\frac{1}{N}\sum_{j=1}^Nq_je^{i\gamma_j}e^{i\theta_j},
\end{equation}
where $Z$ is the mean field whose amplitude $R$ quantifies the extent of the agreement between the oscillators' phases $\theta_i$, while $Y$ represents the weighted mean field, with amplitude $W$ reflecting the agreement between $\theta_i+\gamma_i$ $(+\pi\mbox{ for }q_i<0)$, weighted by $|q_i|$. The model (\ref{km}) can then be rewritten in terms of $Y$ as
\begin{equation}\label{kmw}
\dot{\theta_i}=\omega_i-k_iW\sin(\theta_i-\beta_i-\Phi),
\end{equation}

\noindent In the continuum limit $N\rightarrow\infty$, the system (\ref{km}) is treated using the probability density function $f(\theta,\Gamma,t)$, representing the probability that the oscillator has parameters $\Gamma$ and phase $\theta$ at time $t$. It can be further factorized as
\begin{equation}\label{rho}
f(\theta,\Gamma,t)\equiv\rho(\theta,t|\Gamma)G(\Gamma)
\end{equation}
where $\rho(\theta,t|\Gamma)$ is the conditional probability density function (CPDF), reflecting the probability that at time $t$ the oscillator has a phase $\theta$, given its parameters $\Gamma$. By definition, the CPDF should satisfy $\int_{-\pi}^\pi \rho(\theta,t|\Gamma)d\theta=1$, which leads to the continuity equation
\begin{equation}\label{ceq}
\partial_t\rho(\theta,t|\Gamma)+\partial_\theta\big\{[\omega-kW\sin(\theta-\beta-\Phi)]\rho(\theta,t|\Gamma)\big\}=0,
\end{equation}
where we have used expression (\ref{kmw}) for $\dot{\theta}$.

The CPDF can usually \cite{Pikovsky:08,Pikovsky:11,Ott:09,Ott:11} be represented by the ansatz introduced by Ott and Antonsen (OA ansatz) \cite{Ott:08}
\begin{equation}\label{oa}
\begin{aligned}
&\rho(\theta,t|\Gamma)=\frac{1}{2\pi}\Big[1+2\,{\operatorname{Re}}\frac{a(\Gamma,t) e^{-i\theta}}{1-a(\Gamma,t) e^{-i\theta}}\Big],\;|a(\Gamma,t)|\leq1\\
&\Rightarrow \int_{-\pi}^\pi e^{i\theta}\rho(\theta,t|\Gamma)=a(\Gamma,t)
\end{aligned}
\end{equation}
Note that the definition of $a(\Gamma,t)$ differs from the original OA-variable $\alpha(\Gamma,t)$, as introduced in \cite{Ott:08}, being its complex conjugate: $a=\alpha^*$.

Substituting (\ref{oa}) into (\ref{ceq}) and (\ref{ZYdef}), one obtains
\begin{eqnarray}
&&\frac{\del a}{\del t}-i\omega a+\frac{kW}{2}(a^2e^{-i(\beta+\Phi)}-e^{i(\beta+\Phi)})=0, \label{oaeq}\\
&&Y=We^{i\Phi}=\int qe^{i\gamma}a(\Gamma,t)G(\Gamma)d\Gamma, \label{Y}\\
&&Z=Re^{i\Psi}=\int a(\Gamma,t)G(\Gamma)d\Gamma, \label{Z}
\end{eqnarray}
where here and below, unless specified, the integration over $d\Gamma\equiv dkdq d\beta d\gamma$ is taken over the whole domain ($(-\infty,\infty)$ for $\omega,k,q$ and $[-\pi,\pi]$ for $\beta,\gamma$). Note that, unlike $\Phi$ and $\Psi$ individually, the difference $\Phi-\Psi$ is invariant under the phase shifts $\theta\rightarrow\theta-\varphi(t)$, and represents another meaningful parameter.

\section*{Parameter redundancy}

Parametrizing KM (\ref{km}) with both $\beta_i$ and $\gamma_j$ is mathematically redundant, since one of them can be removed by a change of variables. Thus, in terms of $\tilde{\theta}_i=\theta_i-\beta_i$ the system (\ref{km}) becomes
\begin{equation}\label{kmtr}
\dot{\tilde{\theta}}_i=\omega_i-\frac{k_i}{N}\sum_j q_j\sin(\tilde{\theta}_i-\tilde{\theta}_j-\tilde{\gamma}_j),\;\tilde{\gamma}_i=\gamma_i+\beta_i
\end{equation}
with the new distribution of parameters $\widetilde{G}(\tilde{\Gamma})$ being given as $\widetilde{G}(\omega,k,q,\tilde{\gamma})=\int G(\omega,k,q,\beta,\gamma)\delta(\tilde{\gamma}-\beta-\gamma)d\beta d\gamma$. Taking into account the relationship between the variables in (\ref{kmtr}) and (\ref{km}), it is clear that the conditional distributions of the new phases $\widetilde{\rho}(\tilde{\theta},t|\omega,k,q,\tilde{\gamma})$ are related to the original ones by $\rho(\theta,t|\omega,k,q,\beta,\gamma)=\widetilde{\rho}(\theta-\beta,t|\omega,k,q,\beta+\gamma)$. Substituting the latter into (\ref{ZYdef}), it can be shown that the weighted mean fields in terms of $\theta_i$ and $\tilde{\theta}_i$ are equal, $\widetilde{Y}(t)=Y(t)$, but that the ordinary mean fields can in general change in a non-trivial fashion, $\widetilde{Z}(t)\neq Z(t)$.

Nevertheless, in some cases both $\beta_i$ and $\gamma_j$ can have physical meanings, in which case the ``full'' model (\ref{km}) will provide a more straightforward and meaningful description of the system, while (\ref{kmtr}) will represent only a mathematical formalism. Thus, e.g.\ $Z$ might be related to the real physical quantity, such as the total power output of the laser array, while $\widetilde{Z}$ can be purely mathematical characteristics. To preserve generality, and also to avoid complicating the discussion by introducing different variable transformations, we therefore study the model (\ref{km}) with both $\beta$ and $\gamma$. Additionally, as will be seen below, the distributed $\beta$ has nontrivial consequences in terms of the non-equilibrium dynamics.

\section*{Parameter normalization}

The model (\ref{km}) is invariant under $(k_i,\beta_i)\rightarrow(-k_i,\beta_i+\pi)$, or $(q_i,\gamma_i)\rightarrow(-q_i,\gamma_i+\pi)$, or rescalings $(k_i,q_i)\rightarrow (k_i/r,rq_i)$. To avoid ambiguity, one can fix the normalization of $q_i$ and specify a rule for choosing the sign of $k_i,q_i$; but different choices might be preferred in different situations. For example, the normalization $\sum q_i=1$ is inapplicable when $q_i=\cos(\gamma_i)$; $\sum |q_i|=1$, on the other hand, fails for a Lorenzian distribution of $q$. The most universal choice is $\int|J(\omega,k)|d\omega dk=1$, where $J(\omega,k)$ is defined in (\ref{IJ}) below: it is always applicable and assures $W\leq1$. However, when $q=1$ and $\beta$ is distributed (so $Y=Z$), it will lead to a rescaling of $q$ (so $Y\neq Z$), which might be inconvenient. Therefore, because the normalization can be fixed at any time, we will retain this ambiguity in order to preserve generality. Note, that $W$ (but not $R$ and not $|k|W$) changes under the $(k,q)$-rescalings and thus can be higher than unity.

\section*{Terminology and notations}

We adopt terminology similar to that introduced earlier \cite{Iatsenko:13,Petkoski:13}. The KM form (\ref{km}) is invariant under $\theta'=\theta-\Omega t$, which just changes the natural frequencies to $\omega'=\omega-\Omega$, so the parameter distribution becomes $G'(\omega',k,q,\beta,\gamma)=G(\omega'+\Omega,k,q,\beta,\gamma)$. Thus, one can consider the KM in different frames rotating at frequency $\Omega$ with respect to some chosen frame. For the latter we select the frame with zero mean frequency $\langle\omega\rangle\equiv\int \omega G(\Gamma)d\Gamma=0$ and call it the \emph{natural frame}; the distribution $G(\Gamma)$ is defined in this frame. We define a \emph{stationary state (SS)} as a state with time-independent CPDF $\partial_t\rho(\theta,t|\Gamma)=0$, which also implies $\partial_t Z=\partial_t Y=0$. The state can be stationary only in a particular rotating frame, so it is characterized by its frame frequency $\Omega$ and mean fields $Z,Y$. We refer to SSs with $\Omega=0$ as \emph{natural states (NS)}, while SSs with $\Omega\neq0$ are \emph{traveling wave (TW)} states. For later convenience we define
\begin{equation}\label{nt}
\begin{aligned}
&G_{\Omega}^{\pm}(\Gamma)\equiv G(\pm\omega+\Omega,k,q,\beta,\gamma),\;L(x,\Delta)\equiv\frac{\Delta/\pi}{x^2+\Delta^2},\\
&P(\phi,r)\equiv\frac{(1-r^2)/2\pi}{(1+r)^2+4r\sin^2(\phi/2)}=\frac{r^{-1}(1-r^2)e^{i\phi}/2\pi}{(e^{i\phi}-r)(r^{-1}-e^{i\phi})},\\
\end{aligned}
\end{equation}
with $0\leq r\leq 1$; note, that $P(\phi,0)=1/2\pi$ and $P(\phi,1)=\delta(\phi)$.

The distribution $P(\phi,r)$ (\ref{nt}) represents a Poisson kernel for the unit disc and is a very common for the KM. For example, it can be shown that the OA ansatz (\ref{oa}) can be rewritten \cite{Marvel:09} as $\rho(\theta,t|\Gamma)=P(\theta-\operatorname{arg}[a(\Gamma,t)],|a(\Gamma,t)|)$. It has one pole inside and one outside the unit circle of $z_\phi\equiv e^{i\phi}$, thus being very convenient in relation to analytic derivations, e.g.\ one has $\int_{-\pi}^\pi e^{i\phi}P(\phi-\phi_0,r)d\phi=re^{i\phi_0}$. To simulate $P(\phi,r)$, one sets $\phi_i=2\arctan\big[\frac{1-r}{1+r}\tan\big(\pi(p_i-1/2)\big)\big]$, where $p_i\in[0,1]$ are uniformly distributed random numbers.

We also introduce
\begin{equation}\label{IJ}
\begin{aligned}
&I(\omega,k)\equiv\int e^{i\beta}G(\Gamma)dq d\beta d\gamma,\; I_{\Omega}^{\pm}\equiv I(\pm\omega+\Omega,k),\\
&J(\omega,k)\equiv\int qe^{i(\beta+\gamma)}G(\Gamma)dq d\beta d\gamma,\; J_{\Omega}^{\pm}\equiv J(\pm\omega+\Omega,k).\\
\end{aligned}
\end{equation}
As can be seen, $I$ and $J$ represent an ``effective'' complex distribution of $\omega,k$ for the determination of $Z$ and $Y$, respectively.

\section*{Validity of the OA ansatz}

The OA-description of KM dynamics (\ref{oaeq}),(\ref{Y}),(\ref{Z}), which we use extensively here, has been proven \cite{Ott:09,Ott:11} to hold in the asymptotic limit $t\rightarrow\infty$ for a very large class of frequency distributions. Thus, in most cases, one can safely use these equations to study the system's steady state behavior. As the next step, Pikovsky and Rosenblum derived more general equations \cite{Pikovsky:08,Pikovsky:11} and showed that the full dynamics of the model obeys (\ref{oaeq}),(\ref{Y}),(\ref{Z}) only if the initial phase distribution also belongs to the OA-manifold
\begin{equation}\label{oain}
\rho(\theta,0|\Gamma)=P(\theta-\phi_0(\Gamma),r_0(\Gamma))
\end{equation}
with any $\phi_0(\Gamma)$ and $r_0(\Gamma)$. Although not given explicitly by Pikovsky and Rosenblum, the form (\ref{oain}) can be deduced from their Eq.\ (A.3) in \cite{Pikovsky:11}, which generates (\ref{oain}) in the case of a uniform distribution of the constants of motion $\psi_k$ (for which OA-description was proven to hold \cite{Pikovsky:08,Pikovsky:11}); see also the work of Marvel {\it et al} \cite{Marvel:09}. Note, that the OA-ansatz often provides a good approximation to the system dynamics even when (\ref{oain}) is not satisfied (e.g.\ for $\rho(\theta,0|\Gamma)=R(0)\delta(\theta)+(1-R(0))/2\pi$).

It is often useful to analytically continue $a(\Gamma,t)$ into the complex plane over some of $\Gamma$, e.g.\ to consider $\omega$ to be complex. This is required to apply the conventional OA-reduction procedure \cite{Ott:08} (see also Refs.\ \cite{Montbrio:11a,Montbrio:11b,Pazo:11}) which, where possible, allows one to obtain a finite-dimensional system of equations for $Y(t),Z(t)$. However, one should always have $|a(\Gamma,t)|\leq 1$ as, otherwise, the OA-ansatz (\ref{oa}) becomes ill-defined. Hence $a(\Gamma,t)$ can be considered only in the region of $\Gamma$ for which
\begin{align}
|a(\Gamma,0)|\equiv&\Big|\int e^{i\theta}\rho(\theta,0|\Gamma)d\theta\Big|\leq1, \label{cc1}\\
\partial_t|a(\Gamma,t)|\Big|_{|a|=1}=&-\operatorname{Im}(\omega)+(W/2)\times \nonumber \\
&\times\operatorname{Re}\big[k(e^{i\beta}e^{-i\varphi}-e^{-i\beta}e^{i\varphi})\big]\leq 0\label{cc2}
\end{align}
at any $W$ and $\varphi\equiv\operatorname{arg}[a(\Gamma,t)]-\Phi$. The condition (\ref{cc1}) establishes that $|a|\leq 1$ is satisfied at $t=0$, while (\ref{cc2}) then guarantees that it holds at all other times too. Some manifestations of the issues related to (\ref{cc1}) and (\ref{cc2}) can be found in Sec.\ 3.2.0.1 of Ref.\ \cite{Pikovsky:11} and in Refs.\ \cite{Petkoski:12,Montbrio:11a,Montbrio:11b,Pazo:11}. Note that (\ref{cc1}) is always satisfied when there is no correlation of initial phases with the system parameters $\Gamma$, i.e.\ $\rho(\theta,0|\Gamma)=P(\theta-\Psi(0),R(0))$; unless specified, we will assume that system starts from such a configuration.

\section*{Stationary states}

To describe the steady state behavior of the system (\ref{km}), we consider its SSs, treating them in frames where they are stationary. The parameter distribution then becomes $G(\Gamma)\rightarrow G_{\Omega}^+(\Gamma)$, with $\Omega$ denoting SS frame frequency. By definition and (\ref{oa}), SSs satisfy $\partial_t a=0$. Using this in (\ref{oaeq}) and taking account of the OA validity condition $|a|\leq1$, one finds that all possible SSs in their own rotating frames are described by
\begin{equation}\label{oass}
a_s(\Gamma)=e^{i(\Phi+\beta)}\times
\left\{\begin{array}{l}
\frac{\sqrt{k^2W^2-\omega^2}+i\omega}{kW}\mbox{ if }|\omega|\leq|k|W,\\
i\frac{\omega-{\rm sign}(\omega)\sqrt{\omega^2-k^2W^2}}{kW}\mbox{ if }|\omega|>|k|W.\\
\end{array}\right.
\end{equation}
Using (\ref{oass}) in (\ref{oa}), one can also recover the stationary phase distribution
\begin{equation}\label{cpdfss}
\rho_s(\theta|\Gamma)=
\left\{\begin{array}{l}
\begin{array}{r}
\delta(\theta-\beta-\arcsin(\frac{\omega}{|k|W})-\Phi+\pi {\rm H}(-k))\\
\mbox{ if }|\omega|\leq|k|W,\\
\end{array}\\
\frac{\sqrt{\omega^2-k^2W^2}/2\pi}{|\omega-kW\sin(\theta-\beta-\Phi)|}\mbox{ if }|\omega|>|k|W,\\
\end{array}\right.
\end{equation}
where ${\rm H}(\cdot)$ denotes a Heaviside function. Note that, in addition to $a_s(\Gamma)$ (\ref{oass}), there exists one more stationary solution of (\ref{oaeq}) (the same as (\ref{oass}), but with minus before $\sqrt{k^2W^2-\omega^2}$ for $|\omega|<|k|W$); however, it is never realized in reality because it corresponds to the unstable position on the phase circle, as can be seen by recovering the respective CPDF (same as (\ref{cpdfss}), but with $\theta\rightarrow\theta+\pi$ for $|\omega|<|k|W$).

\section*{Self-consistency and stability conditions}

The macroscopic parameters of the SSs can be found from the self-consistency conditions (SCCs). Using (\ref{oass}) in (\ref{Y}), they become
\begin{equation}\label{scc}
\begin{aligned}
\widetilde{F}\equiv&\int\frac{dk}{kW}{\Big\{}\int_{-|k|W}^{|k|W}\sqrt{k^2W^2-\omega^2}J_\Omega^+d\omega+i\int \omega
J_\Omega^+d\omega-\\
&-i\int_{|k|W}^\infty \sqrt{\omega^2-k^2W^2}[J_\Omega^+-J_\Omega^-]d\omega{\Big\}}=W.\\
\end{aligned}
\end{equation}
From its real and imaginary parts the SS parameters $W,\Omega$ can be determined, and subsequently used in (\ref{oass}),(\ref{Z}) to determine $Z$ as well.

An important property of the SS is its stability. For the incoherent state, it can be analysed using the approach of Ref.\ \cite{Strogatz:91}. Thus, performing a linear stability analysis of (\ref{oa}) above incoherence ($a=0$) and using a self-consistency argument (see \cite{Iatsenko:13} and references therein), one can show that incoherence changes stability when there exists a solution $\Omega$ to the equations
\begin{equation}\label{isc}
\left\{\begin{aligned}
&\pi{\rm Im}\int kJ(\Omega,k)dk+{\rm Re}\int\frac{kJ(\omega+\Omega,k)}{\omega}d\omega dk=0,\\
&\pi{\rm Re}\int kJ(\Omega,k)dk-{\rm Im}\int\frac{kJ(\omega+\Omega,k)}{\omega}d\omega dk=2.\\
\end{aligned}\right.
\end{equation}
Note that, if phase shifts in $\gamma$ or $\beta$ are present then, with increasing coupling strength, the incoherence can not only lose, but also gain \cite{Omelchenko:12,Omelchenko:13}, stability at the transition points determined by (\ref{isc}).

Following Ref.\ \cite{Iatsenko:13} (but with $Z\rightarrow Y$), we can also derive the empirical stability conditions (ESCs) for (\ref{km}):
\begin{equation}\label{esc}
\left\{\begin{aligned}
&\partial_W {\rm Re}\widetilde{F}+W(\partial_\Omega {\rm Im}\widetilde{F})<1,\\
&(\partial_W {\rm Re}\widetilde{F}-1)(\partial_\Omega {\rm Im}\widetilde{F})-(\partial_W {\rm Im}\widetilde{F}) (\partial_\Omega {\rm Re}\widetilde{F})>0.\\
\end{aligned}\right.
\end{equation}
which give the approximate stability conditions for SSs with $W>0$. Although empirical, the ESCs (\ref{esc}) work well in the majority of cases, though not in all, e.g.\ they can fail in the presence of standing waves \cite{Iatsenko:13}; their performance also becomes less good when $\operatorname{arg}[J(\omega,k)]\neq0$. However, ESCs seem to be exact if $\operatorname{arg}[J(\omega,k)]=0$ and the distribution $G(\Gamma)$ is unimodal over $\omega$.

\section*{Uncoupled distributions}

All previous expressions simplify greatly if we consider the distribution of $q,\beta,\gamma$ to be uncorrelated with $\omega,k$. In what follows, we will therefore assume
\begin{equation}\label{uc}
G(\omega,k,q,\beta,\gamma)=g(\omega,k)h(q,\beta,\gamma),\;g_\Omega^{\pm}\equiv g(\pm\omega+\Omega,k),
\end{equation}
so that the effective $(\omega,k)$-distributions (\ref{IJ}) become
\begin{equation}\label{IJuc}
\begin{aligned}
&I(\omega,k)=|I|e^{i\phi_I}g(\omega,k),\;J(\omega,k)=|J|e^{i\phi_J}g(\omega,k),\\
&|I|e^{i\phi_I}\equiv \int e^{i\beta}h(q,\beta,\gamma)dq d\beta d\gamma,\\
&|J|e^{i\phi_J}\equiv \int qe^{i(\beta+\gamma)}h(q,\beta,\gamma)dq d\beta d\gamma.\\
\end{aligned}
\end{equation}

\noindent Then the SCCs (\ref{scc}) simplify to
\begin{equation}\label{sccuc}
F_W(W,\Omega)=W\cos\phi_J/|J|,\;F_\Omega(W,\Omega)=-W\sin\phi_J/|J|,
\end{equation}
where
\begin{equation}\label{FWO}
\begin{aligned}
F_W(W,\Omega)\equiv&\int \frac{dk}{kW}\int_{-|k|W}^{|k|W} g_\Omega^+\sqrt{K^2R^2-\omega^2}d\omega,\\
F_\Omega(W,\Omega)\equiv&\int \frac{dk}{kW}{\Big\{}\int\omega g_\Omega^+d\omega\\
&-\underset{|k|W}{\overset{+\infty}{\int}}{\big[}g_\Omega^+-g_\Omega^-{\big]}\sqrt{\omega^2-k^2W^2}d\omega{\Big\}}\\
\end{aligned}
\end{equation}
are fully analogous to $F_{R,\Omega}$ in Ref.\ \cite{Iatsenko:13}.

The incoherence stability condition (\ref{isc}) becomes
\begin{equation}\label{iscuc}
\left\{\begin{aligned}
&\int\frac{kg(\omega+\Omega,k)}{\omega}d\omega dk
=-\frac{2\sin\phi_J}{|J|},\\
&\int kg(\Omega,k)dk=\frac{2\cos\phi_J}{\pi|J|},\\
\end{aligned}\right.
\end{equation}
while the ESCs (\ref{esc}) can be simplified using $\widetilde{F}=|J|e^{i\phi_J}(F_W+iF_\Omega)$. Note that, for uncorrelated $g(\omega,k)=g(\omega)p(k)$, the stability of incoherence is fully determined by $\langle k\rangle\equiv\int kp(k)dk$ and not by the particular form of $p(k)$, as noticed previously for a simpler KM \cite{Iatsenko:13,Montbrio:11b}.

Furthermore, from (\ref{Z}),(\ref{Y}) and (\ref{oass}) it follows that
\begin{equation}\label{Zuc}
R=|I|W/|J|,\;\Phi-\Psi=\phi_J-\phi_I.
\end{equation}

\section*{System reduction}

From (\ref{sccuc}-\ref{Zuc}) it is evident that all the macroscopic properties of the SSs are completely characterized by $g(\omega,k)$, $|J|$ and $\phi_J$, while $|I|,\phi_I$ serve merely to specify $Z$ (\ref{Zuc}). Instead of (\ref{km}), therefore, one can consider the system
\begin{equation}\label{skeq}
\dot{\theta_i}=\omega_i-\frac{k_i|J|}{N}\sum_{j=1}^N\sin(\theta_i-\theta_j-\phi_J)
\end{equation}
with the same distribution of $\omega_i,k_i$ defined by $g(\omega,k)$. Then each SS of (\ref{skeq}) corresponds to an SS of (\ref{km}), and their parameters are related as
\begin{equation}\label{meq}
\Omega=\Omega_{SK},\;W=|J|R_{SK},\;R=|I|R_{SK},
\end{equation}
where subscript $SK$ denotes the states of the model (\ref{skeq}); the stability of the corresponding states will also be the same, as follows from (\ref{iscuc}) for incoherence, and from the ESCs (\ref{esc}) and numerical evidence for other SSs.

\begin{figure}[t!]
\includegraphics[width=1\linewidth]{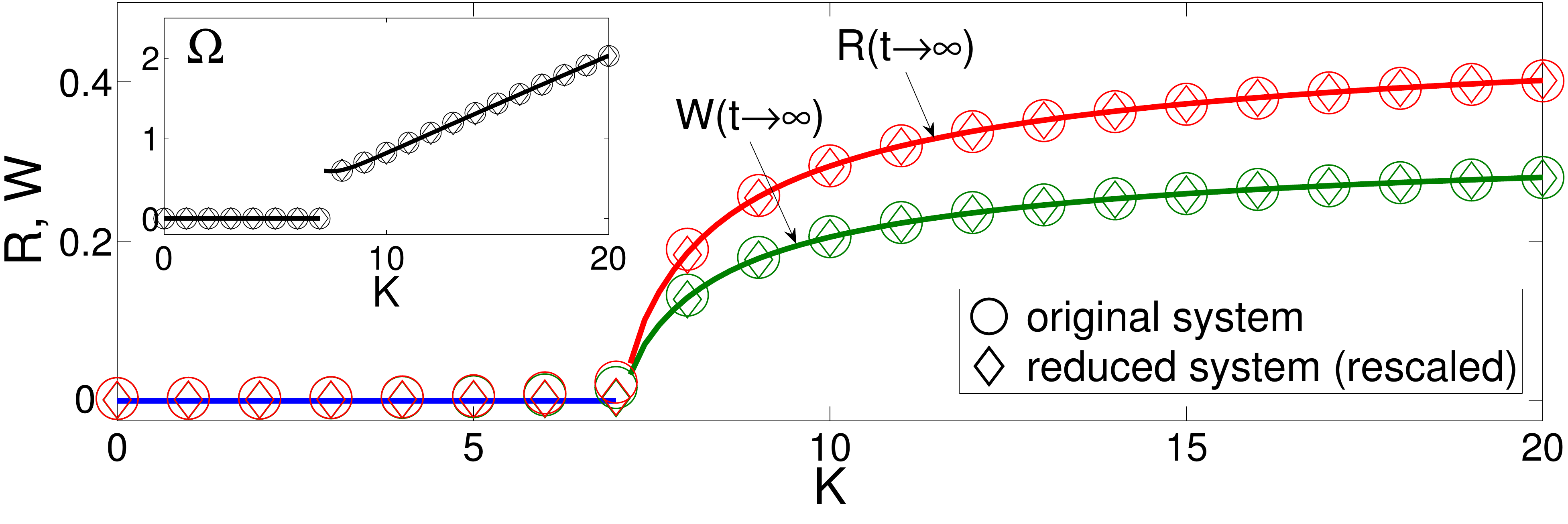}\\
\caption{Dependences of the SS parameters $W$ (green), $R$ (red) and $\Omega$ (black) on coupling $K$, as calculated for the original system (\ref{km}) (circles) and obtained from the reduced system (\ref{skeq}),(\ref{meq}) (diamonds); solid lines show theoretical predictions based on (\ref{sccuc}-\ref{Zuc}). The distribution for the original system was:
$g(\omega,k)\sim\delta(k-K)[\omega^2+e^{-\omega^2}]^{-1},h(q,\beta,\gamma)\sim e^{-(q-1)^2/0.02}[P(\beta-\pi/8,0.8)+P(\beta+\pi/2,0.9)]P(\gamma-\pi/3,0.7)$, but the same picture appears for any $h(q,\beta,\gamma)$ with the same $\phi_J,|J|,|I|$ (\ref{IJuc}). The simulations used $N=10^5$ oscillators and a Runge-Kutta 6th order method with time-step 0.01 s for 500 s; the values are averages over the last 100 s.}
\label{eqfig}
\end{figure}

Hence, for \emph{any} distribution of $q,\beta,\gamma$ obeying (\ref{uc}), one can reduce the consideration of (\ref{km}) to the much simpler Sakaguchi-Kuramoto model (\ref{skeq}). This elegant and unexpected result is illustrated in Fig.\ \ref{eqfig}. Interestingly, something similar was noted earlier \cite{Hong:12} in a less general model than (\ref{km}). The reduction (\ref{meq}), however, relates only to SSs ($t\rightarrow\infty$), whereas the full evolutions $Z(t),Y(t)$ and/or microscopic properties cannot be obtained in this way. Note that, from (\ref{meq}), $W\leq|J|,R\leq|I|$, i.e.\ the distribution of $q,\beta,\gamma$ imposes an upper bound on the SSs' mean field strengths, which they cannot exceed however strong the coupling is.

\section*{Glassy states}

Daido reported evidence \cite{Daido:92} that the KM with random $k_iq_j$ and $\beta_i+\gamma_j$ can undergo a glass transition but, due to the complexity of the model, the resultant ``oscillator glass'' and its properties remained mysterious \cite{Daido:92,Bonilla:93,Stiller:98,Daido:00,Stiller:00}. However, as will be seen below, these exotic states can appear in the model (\ref{km}), in which case they can be studied analytically.

In general, the glassy state can be defined as a state with a uniform distribution of phases $\theta_i$, indistinguishable from incoherence but where, in contrast to the latter, the oscillators adjust their frequencies $\dot{\theta}_i$. The conditions for the glassy behavior can be formulated as
\begin{eqnarray}
&&\rho(\theta,t)\equiv\int\rho(\theta,t|\Gamma)G(\Gamma)d\Gamma=1/2\pi,\label{glass1}\\
&&Q(\langle\dot{\theta}\rangle_t,\omega)>0,\label{glass2}
\end{eqnarray}
where $\langle\cdot\rangle_t$ denotes a time-average, $\Gamma$ here stands for any general parameters of the KM considered (not only that of the form (\ref{km})), and $Q$ quantifies the degree of frequency adjustment between the oscillators as compared to incoherence (for which $\langle\dot{\theta}\rangle_t=\omega$). The latter can be chosen e.g.\ as the difference between the Shannon entropies for the corresponding distributions, $Q(x,y)=\int p_x(x)\log p_x(x)dx-\int p_y(y)\log p_y(y)dy$, where $p_{x,y}$ are the marginal distributions of $x$ and $y$, respectively.

\begin{figure*}[t!]
\includegraphics[width=1\linewidth]{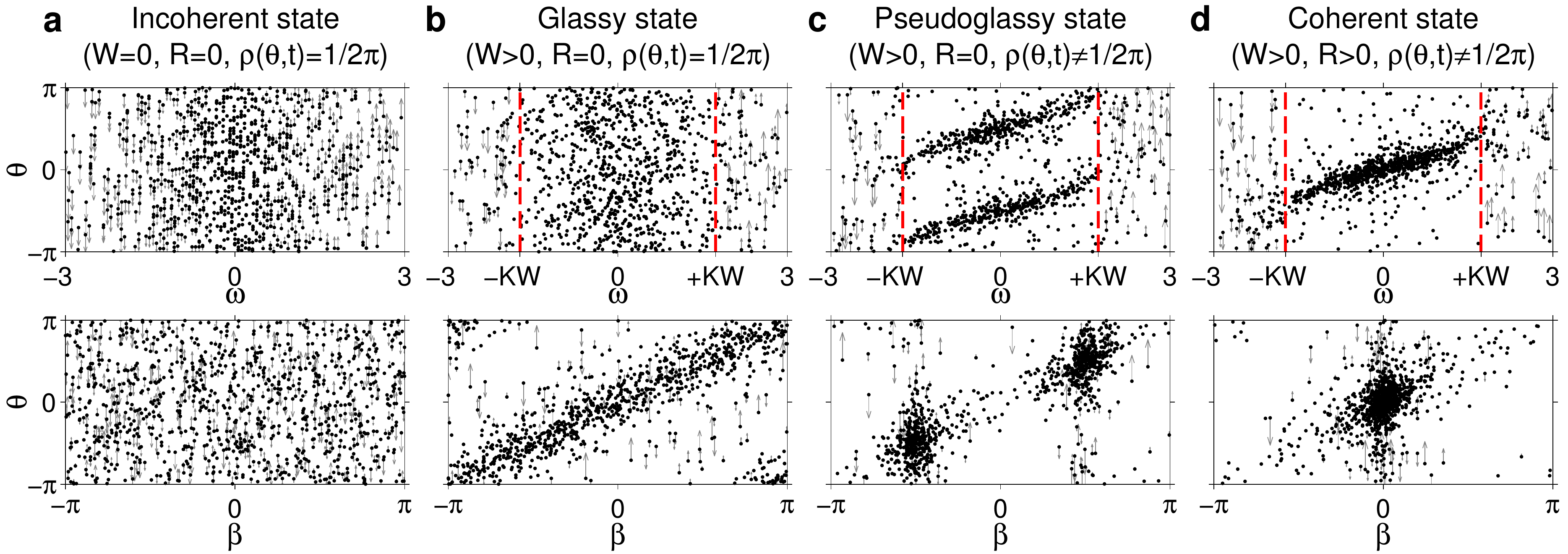}
\caption{The different kinds of states, as illustrated by snapshots of positions in the $\theta,\omega$ (upper panel) and $\theta,\beta$ (bottom panel) planes of (the same) 1000 randomly selected oscillators (out of $N=25600$). Tiny vertical arrows show the oscillators movements during 0.25 s  (for a dynamic version of the figure see Ref.\ \cite{suppmat}).
Red dashed lines show boundaries between the clusters and incoherent populations. In all cases $g(\omega,k)=L(\omega,1)\delta(k-K),\;h_2(q,\gamma|\beta)=\delta(q-1)\delta(\gamma+\beta)$, and (a) $K=1,\;h_1(\beta)=1/2\pi$; (b) $K=3,\;h_1(\beta)=1/2\pi$; (c) $K=3,\;h_1(\beta)=(1/2)[P(\beta-\pi/2,0.8)+P(\beta+\pi/2,0.8)]$; (d) $K=3,\;h_1(\beta)=P(\beta,0.8)$. The pictures will be the same for any $h_2(q,\gamma|\beta)$ satisfying $|J|=1,\phi_J=0$ in each case. All states have $\Omega=0$ and are presented in the natural frame ($\langle\omega\rangle=0$); for $\Omega\neq0$, corresponding to glassy, pseudoglassy or coherent traveling waves, one would observe similar pictures in the rotating frames.}
\label{iglassy}
\end{figure*}

Note that the condition for the absence of phase-locking is sometimes taken as $R=0$. However, although being implied by (\ref{glass1}), it is not as rigorous as the latter. Thus, there might be many configurations for which $R=0$ but where the oscillators adjust their phases. Examples include e.g.\ two phase-locked populations of the same size, which are in anti-phase with each other. States with $R=0$ satisfying (\ref{glass2}), but not (\ref{glass1}), will be called \emph{pseudoglassy}. Note that, according to this classification, the glassy synchronization reported in \cite{Bonilla:93} is in fact pseudoglassy, as the corresponding states consist of the incoherent population and two synchronized clusters of opposite phase. All stationary states with $R>0$, including the usual synchronized states, $\pi$-states \cite{Hong:11a,Hong:11b} and traveling waves, will be classified as \emph{coherent}.

As a simplified picture, the distinctions between the states can be understood in terms of a group of people doing cyclical exercises, each with their own tempo and other parameters. The incoherent state is when everyone proceeds independently; the coherent state is when they all move synchronously, at any given time having similar poses; pseudoglassy is when they exercise at the same tempo but remain pairwise in opposite poses; and glassy is when they adjust their tempos, but always remain in the same independent random poses, thus representing a kind of synchronous disorder.

Considering (\ref{km}) with the uncoupled distribution (\ref{uc}), it can be shown that glassy states can appear if
\begin{align}
&\int h(q,\beta,\gamma)dq d\gamma=1/2\pi,\label{cglassy1}\\
&|J|\equiv\Big|\int qe^{i(\gamma+\beta)}h(q,\beta,\gamma)dq d\beta d\gamma\Big|>0,\label{cglassy2}
\end{align}
in which case all stationary states except incoherence satisfy (\ref{glass1}) and (\ref{glass2}). Thus, as follows from (\ref{cpdfss}), the phases of the oscillators with $|\omega_i|\leq|k_i|W$ are frozen at $\theta_i=\beta_i+\arcsin(\omega_i/|k_i|W)+\pi H(-k_i)$ (all in the rotating frame). For states with $W>0$ an absence of phase-locking between the oscillators (\ref{glass1}) therefore requires a uniform distribution of $\beta_i$ (\ref{cglassy1}). Next, $W>0$ is the necessary and sufficient condition for (\ref{glass2}): necessary because, otherwise, all oscillators have $\langle\theta_i\rangle_t=\omega_i$, as implied by (\ref{kmw}); and sufficient because it establishes frequency-locking of the oscillators with $|\omega_i|\leq|k_i|W$ (as their phases are constant). Finally, as is clear from (\ref{sccuc}), SSs with $W>0$ can appear only if (\ref{cglassy2}) is satisfied. Note, that the exact conditions (\ref{cglassy1}), (\ref{cglassy2}) for the appearance of the glassy states have been found for the first time.

Representing $h(q,\beta,\gamma)\equiv h_1(\beta)h_2(q,\gamma|\beta)$, the condition (\ref{cglassy2}) is equivalent to $h_1(\beta)=1/2\pi$. Thus to satisfy also (\ref{cglassy2}), the distribution of $q,\gamma$ should be specifically correlated with $\beta$. The simplest examples are $h(q,\gamma|\beta)=\delta(q-q_0)P(\gamma+\beta,r)$ and $h(q,\gamma|\beta)=L(q-q_0\cos\beta,\Delta)P(\gamma-\gamma_0,r)$. For pseudoglassy states, the condition (\ref{cglassy1}) is not satisfied, but $R=0$ which, in the present case, is equivalent to $|I|\equiv|\int e^{i\beta}h_1(\beta)d\beta|=0$, as can be seen from (\ref{Zuc}). There are many possible $h_1(\beta)$ satisfying the latter, e.g.\ $h_1(\beta)=P(\beta-\beta_0,r)+P(\beta-\beta_0-\pi,r)$.

Examples of the four types of states occurring in the model (\ref{km}) are shown in Figure \ref{iglassy}, presented in the natural frame, where the population does not move as a whole ($\langle\omega\rangle=0$). In the $(\theta,\omega)$-plane the only difference between the glassy state and incoherence is that, in the former, phases within the glassy cluster ($|\omega|<|k|W$) are ``frozen'' around random angles $\beta$ (static disorder), as seen in $(\theta,\beta)$-plane; this is in contrast to their asynchronous movement (dynamic disorder), observed for incoherence.

As discussed previously, the model (\ref{km}) can be reduced to a form without $\beta$ (\ref{kmtr}) by a change of variables. So any glassy and pseudoglassy state in terms of $\theta_i$ (\ref{km}) will correspond to the coherent state in terms of $\tilde{\theta}_i=\theta_i-\beta_i$. Thus, in the model (\ref{km}) the glassy and pseudoglassy states are in a strict mathematical sense redundant. In practice, however, the KM with distributed $\beta$ can provide a more straightforward and physically meaningful description of the system, so these states are realizable.

\section*{Relaxation dynamics}

Up to now, we have concentrated mainly on the behavior of the system (\ref{km}) in the asymptotic limit $t\rightarrow\infty$, represented by a set of its possible SSs. For particular $G(\Gamma)$, one can also obtain the full time-evolutions $Y(t),Z(t)$ by using the OA-reduction procedure \cite{Ott:08}, which can be extended to incorporate the distributed phase shifts. Consider the distribution
\begin{equation}\label{dgamma}
G(\Gamma)=L(\omega,\Delta)\delta(k-K)\delta(q-1)\delta(\beta)P(\gamma-\phi_0,r_0)
\end{equation}
In this case $a(\Gamma,t)$ can be analytically continued inside the unit circle of $z_\gamma=e^{i\gamma}$, because (\ref{cc1}) is satisfied for all $z_\gamma$. Then in (\ref{Y}) and (\ref{Z}) one can  integrate over $\gamma$ by changing it to the unit circle of $e^{i\gamma}$, and then taking the residue at the pole $z_\gamma=r_0e^{i\phi_0}$ of $P(\gamma-\phi_0,r_0)$ (\ref{nt}). The integration over $\omega$ is performed in the usual way \cite{Ott:08}, i.e.\ by taking the residue at $\omega=i\Delta$. Then from (\ref{Y}) and (\ref{Z}) one obtains $Z(t)=a(i\Delta,e^{i\gamma}=r_0e^{i\phi_0},t),Y(t)=r_0 e^{i\phi_0}Z(t)$. Substituting this into (\ref{oa}) and solving the resultant equations, one gets finally
\begin{equation}\label{tevol}
\begin{aligned}
&W(t)=\frac{|J|\sqrt{|1-\frac{2\Delta}{K_{J}}|}}{\sqrt{|1+[(1-\frac{2\Delta}{K_{J}})(|J|/W(0))^2-1]e^{-(K_{J}-2\Delta)t}|}}\\
&\dot{\Phi}(t)\equiv\Omega(t)=\frac{K_{J}\tan\phi_J}{2}\Big[\frac{W^2(t)}{|J|^2}+1\Big],\;K_{J}\equiv K|J|\cos\phi_J\\
&R(t)=|I|W(t)/|J|,\;\Phi(t)-\Psi(t)=\phi_J-\phi_I,\\
\end{aligned}
\end{equation}
where we have represented all in terms of $I,J$ (\ref{IJuc}), which in the present case are $|I|e^{i\phi_I}=1$, $|J|e^{i\phi_J}=r_0e^{i\phi_0}$. Obviously, a similar procedure to that outlined here can be applied if the distribution of $\gamma$ has a few poles inside the unit circle.

Now consider the KM (\ref{km}) with distributed $\beta$:
\begin{equation}\label{dbeta}
G(\Gamma)=L(\omega,\Delta)\delta(k-K)\delta(q-1)P(\beta-\phi_0,r_0)\delta(\gamma),
\end{equation}
in which case one has $Z(t)=Y(t)$ by the definition (\ref{ZYdef}). By the change of variables $\tilde{\theta}_i=\theta_i-\beta_i$ the present case can be reduced to the already-studied KM with distributed $\gamma$ (\ref{dgamma}). Then, since $Z(t)=Y(t)=\widetilde{Y}(t)$ (see discussion of (\ref{kmtr})), the macroscopic parameters should still evolve according to (\ref{tevol}), although now one should use $|I|e^{i\phi_I}=|J|e^{i\phi_J}=r_0e^{i\phi_0}$ in the latter. The same result may be obtained by applying the OA-reduction, with the integrals (\ref{Y}) and (\ref{Z}) over $\beta$ being evaluated by taking residues inside the unit circle of $e^{i\beta}$, similarly to what was done in the previous case.

\begin{figure}[t!]
\includegraphics[width=1\linewidth]{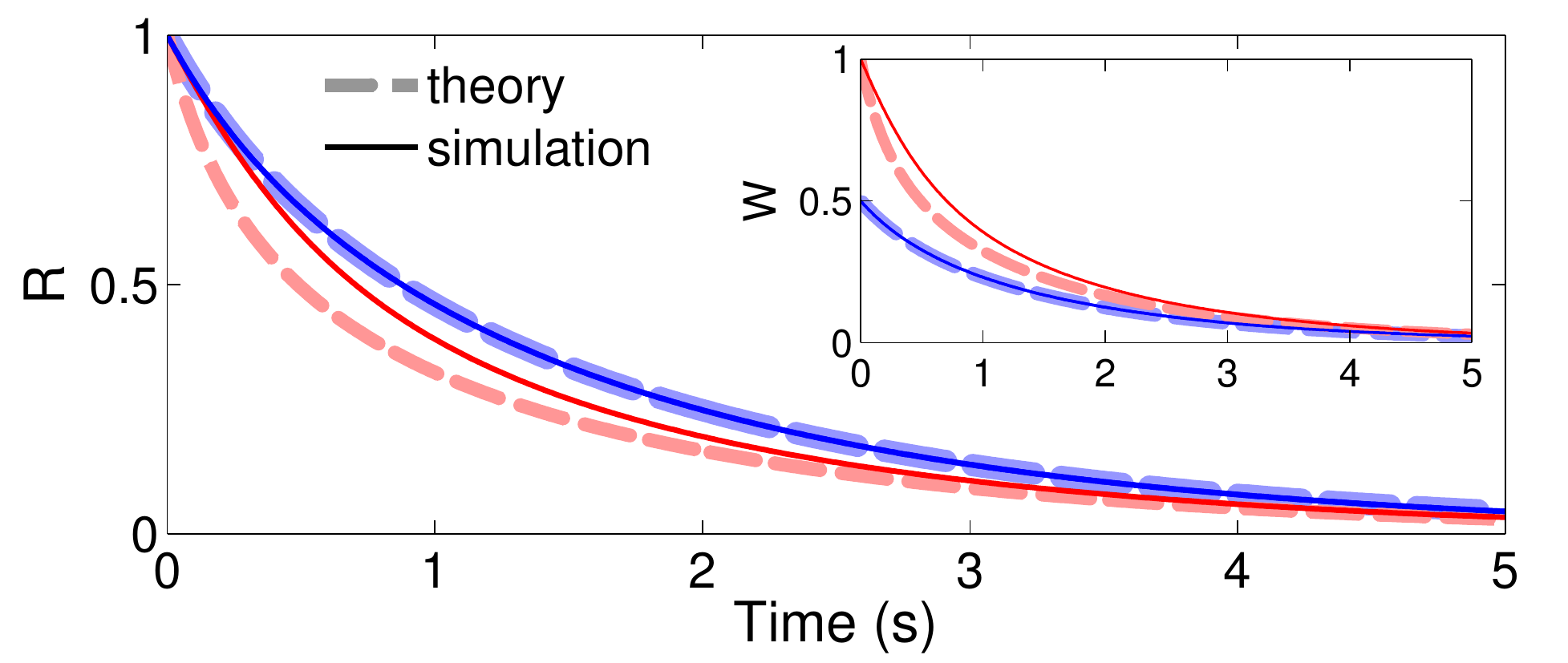}\\
\caption{Time-evolution of $R(t)$ and $W(t)$ (inset) from the initial conditions $R(0)=1$: blue -- distributed $\gamma$ (\ref{dgamma}), red -- distributed $\beta$ (\ref{dbeta}). Dashed lines show the behavior predicted from (\ref{tevol}), while solid lines correspond to results of numerical simulations. In (\ref{dgamma}) and (\ref{dbeta}) we used $\phi_0=\pi/6,r_0=0.5$.  The simulations used $N=10^6$ oscillators and a Runge-Kutta 6th order method with time-step $0.01$ s.}
\label{fig:gammabeta}
\end{figure}

However, the behavior predicted by (\ref{tevol}) and the actual behavior of $Z(t),Y(t)$ agree only for distributed $\gamma$, but not for distributed $\beta$, as demonstrated in Fig.\ \ref{fig:gammabeta}. The reason for this is that, first, one cannot continue $a(\Gamma,t)$ inside the unit circle of $\beta$, as (\ref{cc2}) is not satisfied there, which makes OA-reduction  impossible when $\beta$ is distributed. Secondly, the evolution of $Y(t),Z(t)$ for (\ref{dbeta}), at least starting from initial conditions uncorrelated with $\beta$, cannot be obtained from the system of $\tilde{\theta}_i=\theta_i-\beta_i$. This is because such a transformation changes the initial conditions, introducing their correlation with system parameters. Thus, for $\theta_i=0$ one will have $\tilde{\theta}_i=-\gamma_i$, so that $\widetilde{a}(\Gamma,0)=e^{-i\gamma}$ and the OA reduction cannot be applied because now (\ref{cc1}) is not satisfied. Therefore, although the cases (\ref{dgamma}) and (\ref{dbeta}) can be obtained from each other by change of variables, the time-evolution for distributed $\beta$ appears to be much more complex than for distributed $\gamma$, and cannot be obtained in a simple form.

\section*{Superrelaxation}

\begin{figure}[t!]
\includegraphics[width=1\linewidth]{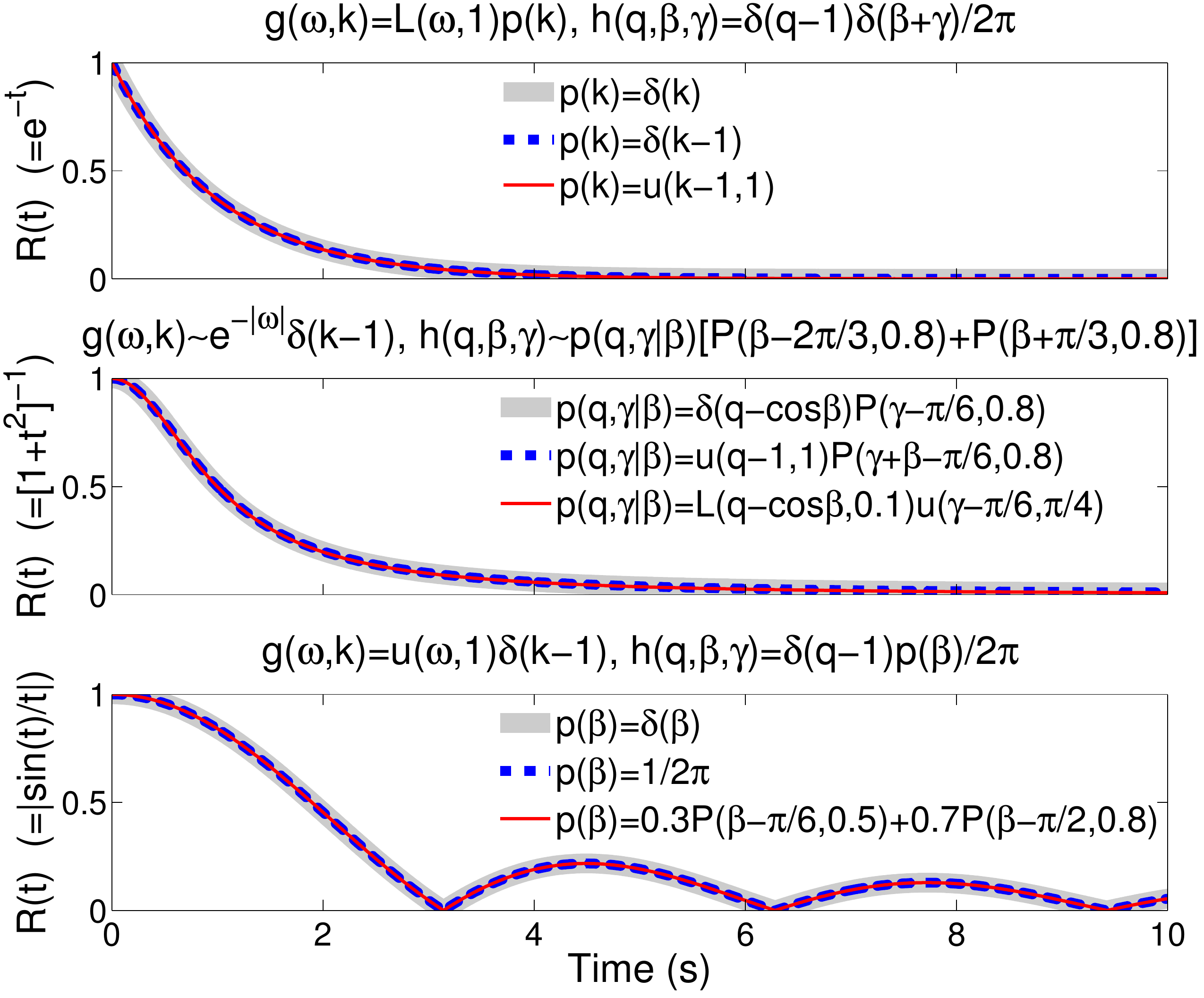}\\
\caption{Interaction-independent relaxation of the order parameter to incoherence for different parameter distributions (all satisfying (\ref{superc})), which are indicated on the figure; $u(x,b)$ denotes uniform distribution of $x$ in $[-b,b]$, while $L(x,\Delta)$ and $P(x,r)$ are defined in (\ref{nt}). The simulations used $N=10^6$ oscillators and a Runge-Kutta 6th order method with time-step $0.01$ s.}
\label{fig:srelax}
\end{figure}

Astonishingly, for a class of distributions $h(q,\beta,\gamma)$ the oscillators do not feel any interaction at all while relaxing from $R(0)=1$ to incoherence, a phenomenon reminiscent of superfluidity/superconductivity. This behavior occurs when
\begin{equation}\label{superc}
\int qe^{i\gamma}h(q,\beta,\gamma)dq d\beta d\gamma=0
\end{equation}
which, unless the initial phases are specifically correlated with system parameters, implies $W(0)=0$. Based on numerical evidence, the weighted mean field then stays at zero during the whole evolution $W(\forall t)=0$, leading to an effective disappearance of interaction between the oscillators, as implied by (\ref{kmw}). As a result, the oscillators evolve freely ($\dot{\theta_i}=\omega_i$) and so the relaxation depends only on the marginal distribution of $\omega$:
\begin{equation}\label{superr}
R(t)=\Big|\int e^{i\omega t}g(\omega,k)d\omega dk\Big|.
\end{equation}
This phenomenon, which we refer to as \emph{superrelaxation}, is illustrated in Fig.\ \ref{fig:srelax}.

\begin{figure}[t!]
\includegraphics[width=1\linewidth]{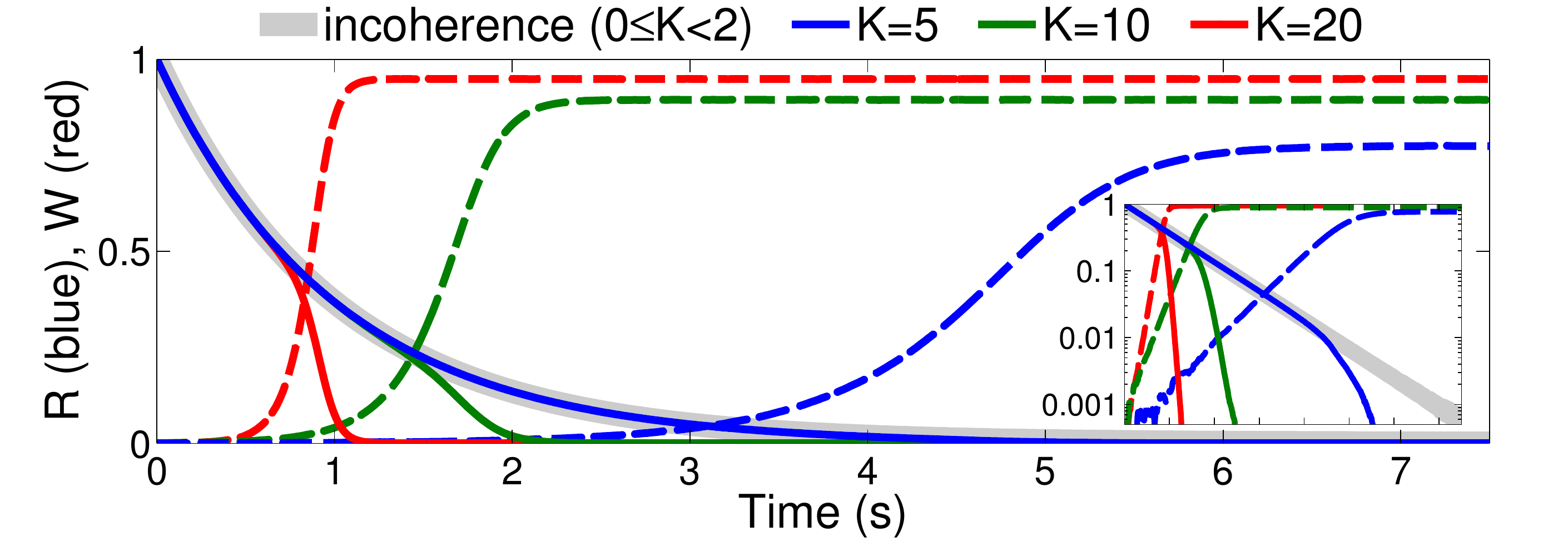}\\
\caption{Time-evolution of $R(t)$ (solid) and $W(t)$ (dashed) from the initial conditions $R(0)=1$ for the glassy states ($g(\omega,k)=L(\omega,1)\delta(k-K),\;h(q,\beta,\gamma)=\delta(q-1)\delta(\beta+\gamma)/2\pi$) for different constant couplings $k_i=K$. The inset shows the results on a logarithmic ordinate scale. The simulations used $N=10^6$ oscillators and a Runge-Kutta 6th order method with time-step $0.01$ s.}
\label{fig:evol}
\end{figure}

Superrelaxation can appear only if the final state is incoherence ($W=0$), while relaxation to SSs with $W>0$ will generally be coupling-dependent, even if (\ref{superc}) is fulfilled. In the latter case it occurs in two stages, as shown in Fig.\ \ref{fig:evol} for the example of glassy states. The phases first begin to disorder in the same way as for incoherence, but are slowly entrained while passing their equilibrium positions. When the field of the entrained oscillators, characterized by $W,$ becomes strong enough, they begin to force unentrained ones to take their positions, so the relaxation switches to a faster, coupling-dependent regime. This ``switch'' occurs sooner for stronger coupling.

\section*{Discussion}

We have generalized our earlier approach \cite{Iatsenko:13} to make it applicable to the more general KM (\ref{km}) so that, using (\ref{scc}), (\ref{isc}) and (\ref{esc}), one can immediately obtain the macroscopic parameters of possible stationary states. The generalized KM (\ref{km}) encompasses a variety of KM variants studied earlier \cite{Iatsenko:13,Hong:11a,Hong:12,Sakaguchi:86,Montbrio:11a,Pazo:11,Montbrio:11b}, allowing one readily to reproduce and extend many of the previous results. Remarkably, the steady state behavior of (\ref{km}) with any distribution $h(q,\beta,\gamma)$ (\ref{uc}) can be obtained from the simple Sakaguchi-Kuramoto model (\ref{skeq}),(\ref{meq}). Note, that all formulas except (\ref{tevol}) can straightforwardly be extended to the case when parameter distribution depends additionally on $W$ (but not $R$ if they differ): $G(\Gamma)\rightarrow G(\Gamma,W)$, which might occur e.g.\ due to nonlinearity (see Refs.\ \cite{Pikovsky:09,Filatrella:07,Rosenblum:07,Baibolatov:10} for motivation and examples). In this case one simply adds this dependence everywhere, setting $J(\omega,k)\rightarrow J(\omega,k,W)\equiv \int qe^{i(\gamma+\beta)}G(\Gamma,W)d\Gamma$ in (\ref{scc}), (\ref{isc}), (\ref{esc}), and similarly for other expressions.

Most interestingly, we have found, that the model (\ref{km}) can exhibit exotic behavior, such as glassy states and superrelaxation, thus opening new horizons for KM-related investigations, both theoretical and practical. These discoveries have a far-reaching implications. For example, it should now be possible to create, observe and study oscillator glass in real systems, where a variety of novel phenomena may be anticipated. As one possible application, if some physical quantity can be associated with the weighted mean field $Y$ (\ref{ZYdef}) in laser arrays \cite{Oliva:01} described by the KM, it might be possible to construct a laser exhibiting zero intensity ($R=0$) but for which other effects are nonvanishing ($W>0$); similar considerations apply to a wide range of different KM applications. Furthermore, the phenomenon of superrelaxation might be used to design a systems whose dynamics remains highly stable in the face of different perturbations and parameter changes.

\begin{acknowledgments}
We are grateful to S.\ Petkoski for valuable discussions of the manuscript. This work was supported by the Engineering and Physical Sciences Research Council (UK) [grant number EP/100999X1].
\end{acknowledgments}

\bibliography{KMbib}

\begin{thebibliography}{43}
\expandafter\ifx\csname natexlab\endcsname\relax\def\natexlab#1{#1}\fi
\expandafter\ifx\csname bibnamefont\endcsname\relax
  \def\bibnamefont#1{#1}\fi
\expandafter\ifx\csname bibfnamefont\endcsname\relax
  \def\bibfnamefont#1{#1}\fi
\expandafter\ifx\csname citenamefont\endcsname\relax
  \def\citenamefont#1{#1}\fi
\expandafter\ifx\csname url\endcsname\relax
  \def\url#1{\texttt{#1}}\fi
\expandafter\ifx\csname urlprefix\endcsname\relax\def\urlprefix{URL }\fi
\providecommand{\bibinfo}[2]{#2}
\providecommand{\eprint}[2][]{\url{#2}}

\bibitem[{\citenamefont{Kuramoto}(1984)}]{Kuramoto:84}
\bibinfo{author}{\bibfnamefont{Y.}~\bibnamefont{Kuramoto}},
  \emph{\bibinfo{title}{Chemical Oscillations, Waves, and Turbulence}}
  (\bibinfo{publisher}{Springer-Verlag}, \bibinfo{address}{Berlin},
  \bibinfo{year}{1984}).

\bibitem[{\citenamefont{Strogatz}(2000)}]{Strogatz:00}
\bibinfo{author}{\bibfnamefont{S.~H.} \bibnamefont{Strogatz}},
  \bibinfo{journal}{Physica D} \textbf{\bibinfo{volume}{143}},
  \bibinfo{pages}{1} (\bibinfo{year}{2000}).

\bibitem[{\citenamefont{Acebr{\'o}n et~al.}(2005)\citenamefont{Acebr{\'o}n,
  Bonilla, Vicente, Ritort, and Spigler}}]{Acebron:05}
\bibinfo{author}{\bibfnamefont{J.~A.} \bibnamefont{Acebr{\'o}n}},
  \bibinfo{author}{\bibfnamefont{L.~L.} \bibnamefont{Bonilla}},
  \bibinfo{author}{\bibfnamefont{C.~J.~P.} \bibnamefont{Vicente}},
  \bibinfo{author}{\bibfnamefont{F.}~\bibnamefont{Ritort}}, \bibnamefont{and}
  \bibinfo{author}{\bibfnamefont{R.}~\bibnamefont{Spigler}},
  \bibinfo{journal}{Rev. Mod. Phys.} \textbf{\bibinfo{volume}{77}},
  \bibinfo{pages}{137} (\bibinfo{year}{2005}).

\bibitem[{\citenamefont{Javaloyes et~al.}(2008)\citenamefont{Javaloyes, Perrin,
  and Politi}}]{Javaloyes:08}
\bibinfo{author}{\bibfnamefont{J.}~\bibnamefont{Javaloyes}},
  \bibinfo{author}{\bibfnamefont{M.}~\bibnamefont{Perrin}}, \bibnamefont{and}
  \bibinfo{author}{\bibfnamefont{A.}~\bibnamefont{Politi}},
  \bibinfo{journal}{Phys. Rev. E} \textbf{\bibinfo{volume}{78}},
  \bibinfo{pages}{011108} (\bibinfo{year}{2008}).

\bibitem[{\citenamefont{Oliva and Strogatz}(2001)}]{Oliva:01}
\bibinfo{author}{\bibfnamefont{R.~A.} \bibnamefont{Oliva}} \bibnamefont{and}
  \bibinfo{author}{\bibfnamefont{S.~H.} \bibnamefont{Strogatz}},
  \bibinfo{journal}{Int. J. Bifurcation Chaos} \textbf{\bibinfo{volume}{11}},
  \bibinfo{pages}{2359} (\bibinfo{year}{2001}).

\bibitem[{\citenamefont{Sheeba et~al.}(2008)\citenamefont{Sheeba, Stefanovska,
  and McClintock}}]{Sheeba:08a}
\bibinfo{author}{\bibfnamefont{J.~H.} \bibnamefont{Sheeba}},
  \bibinfo{author}{\bibfnamefont{A.}~\bibnamefont{Stefanovska}},
  \bibnamefont{and} \bibinfo{author}{\bibfnamefont{P.~V.~E.}
  \bibnamefont{McClintock}}, \bibinfo{journal}{Biophys.\ J.}
  \textbf{\bibinfo{volume}{95}}, \bibinfo{pages}{2722} (\bibinfo{year}{2008}).

\bibitem[{\citenamefont{Wiesenfeld et~al.}(1998)\citenamefont{Wiesenfeld,
  Colet, and Strogatz}}]{Wiesenfeld:98}
\bibinfo{author}{\bibfnamefont{K.}~\bibnamefont{Wiesenfeld}},
  \bibinfo{author}{\bibfnamefont{P.}~\bibnamefont{Colet}}, \bibnamefont{and}
  \bibinfo{author}{\bibfnamefont{S.~H.} \bibnamefont{Strogatz}},
  \bibinfo{journal}{Phys. Rev. E} \textbf{\bibinfo{volume}{57}},
  \bibinfo{pages}{1563} (\bibinfo{year}{1998}).

\bibitem[{\citenamefont{N\'{e}da et~al.}(2000)\citenamefont{N\'{e}da, Ravasz,
  Brechet, Vicsek, and Barabasi}}]{Neda:00}
\bibinfo{author}{\bibfnamefont{Z.}~\bibnamefont{N\'{e}da}},
  \bibinfo{author}{\bibfnamefont{E.}~\bibnamefont{Ravasz}},
  \bibinfo{author}{\bibfnamefont{Y.}~\bibnamefont{Brechet}},
  \bibinfo{author}{\bibfnamefont{T.}~\bibnamefont{Vicsek}}, \bibnamefont{and}
  \bibinfo{author}{\bibfnamefont{A.~L.} \bibnamefont{Barabasi}},
  \bibinfo{journal}{Nature} \textbf{\bibinfo{volume}{403}},
  \bibinfo{pages}{849} (\bibinfo{year}{2000}).

\bibitem[{\citenamefont{Iatsenko et~al.}(2013)\citenamefont{Iatsenko, Petkoski,
  McClintock, and Stefanovska}}]{Iatsenko:13}
\bibinfo{author}{\bibfnamefont{D.}~\bibnamefont{Iatsenko}},
  \bibinfo{author}{\bibfnamefont{S.}~\bibnamefont{Petkoski}},
  \bibinfo{author}{\bibfnamefont{P.~V.~E.} \bibnamefont{McClintock}},
  \bibnamefont{and}
  \bibinfo{author}{\bibfnamefont{A.}~\bibnamefont{Stefanovska}},
  \bibinfo{journal}{Phys. Rev. Lett.} \textbf{\bibinfo{volume}{110}},
  \bibinfo{pages}{064101} (\bibinfo{year}{2013}),
  \urlprefix\url{http://link.aps.org/doi/10.1103/PhysRevLett.110.064101}.

\bibitem[{\citenamefont{Petkoski and Stefanovska}(2012)}]{Petkoski:12}
\bibinfo{author}{\bibfnamefont{S.}~\bibnamefont{Petkoski}} \bibnamefont{and}
  \bibinfo{author}{\bibfnamefont{A.}~\bibnamefont{Stefanovska}},
  \bibinfo{journal}{Phys. Rev. E} \textbf{\bibinfo{volume}{86}},
  \bibinfo{pages}{046212} (\bibinfo{year}{2012}).

\bibitem[{\citenamefont{Montbrio et~al.}(2006)\citenamefont{Montbrio, Pazo, and
  Schmidt}}]{Montbrio:06}
\bibinfo{author}{\bibfnamefont{E.}~\bibnamefont{Montbrio}},
  \bibinfo{author}{\bibfnamefont{D.}~\bibnamefont{Pazo}}, \bibnamefont{and}
  \bibinfo{author}{\bibfnamefont{J.}~\bibnamefont{Schmidt}},
  \bibinfo{journal}{Phys. Rev. E} \textbf{\bibinfo{volume}{74}},
  \bibinfo{pages}{056201} (\bibinfo{year}{2006}).

\bibitem[{\citenamefont{Montbrio and Pazo}(2011{\natexlab{a}})}]{Montbrio:11a}
\bibinfo{author}{\bibfnamefont{E.}~\bibnamefont{Montbrio}} \bibnamefont{and}
  \bibinfo{author}{\bibfnamefont{D.}~\bibnamefont{Pazo}},
  \bibinfo{journal}{Phys. Rev. Lett.} \textbf{\bibinfo{volume}{106}},
  \bibinfo{pages}{254101} (\bibinfo{year}{2011}{\natexlab{a}}).

\bibitem[{\citenamefont{Pazo and Montbrio}(2011)}]{Pazo:11}
\bibinfo{author}{\bibfnamefont{D.}~\bibnamefont{Pazo}} \bibnamefont{and}
  \bibinfo{author}{\bibfnamefont{E.}~\bibnamefont{Montbrio}},
  \bibinfo{journal}{Europhys. Lett.} \textbf{\bibinfo{volume}{95}},
  \bibinfo{pages}{60007} (\bibinfo{year}{2011}).

\bibitem[{\citenamefont{Montbrio and Pazo}(2011{\natexlab{b}})}]{Montbrio:11b}
\bibinfo{author}{\bibfnamefont{E.}~\bibnamefont{Montbrio}} \bibnamefont{and}
  \bibinfo{author}{\bibfnamefont{D.}~\bibnamefont{Pazo}},
  \bibinfo{journal}{Phys. Rev. E} \textbf{\bibinfo{volume}{84}},
  \bibinfo{pages}{046206} (\bibinfo{year}{2011}{\natexlab{b}}).

\bibitem[{\citenamefont{Skardal and Restrepo}(2012)}]{Skardal:12}
\bibinfo{author}{\bibfnamefont{P.~S.} \bibnamefont{Skardal}} \bibnamefont{and}
  \bibinfo{author}{\bibfnamefont{J.~G.} \bibnamefont{Restrepo}},
  \bibinfo{journal}{Phys. Rev. E} \textbf{\bibinfo{volume}{85}},
  \bibinfo{pages}{016208} (\bibinfo{year}{2012}).

\bibitem[{\citenamefont{Anderson et~al.}(2012)\citenamefont{Anderson, Tenzer,
  Barlev, Girvan, Antonsen, and Ott}}]{Anderson:12}
\bibinfo{author}{\bibfnamefont{D.}~\bibnamefont{Anderson}},
  \bibinfo{author}{\bibfnamefont{A.}~\bibnamefont{Tenzer}},
  \bibinfo{author}{\bibfnamefont{G.}~\bibnamefont{Barlev}},
  \bibinfo{author}{\bibfnamefont{M.}~\bibnamefont{Girvan}},
  \bibinfo{author}{\bibfnamefont{T.~M.} \bibnamefont{Antonsen}},
  \bibnamefont{and} \bibinfo{author}{\bibfnamefont{E.}~\bibnamefont{Ott}},
  \bibinfo{journal}{Chaos} \textbf{\bibinfo{volume}{22}},
  \bibinfo{pages}{013102} (\bibinfo{year}{2012}).

\bibitem[{\citenamefont{Lee et~al.}(2011)\citenamefont{Lee, Restrepo, Ott, and
  Antonsen}}]{Lee:11}
\bibinfo{author}{\bibfnamefont{W.~S.} \bibnamefont{Lee}},
  \bibinfo{author}{\bibfnamefont{J.~G.} \bibnamefont{Restrepo}},
  \bibinfo{author}{\bibfnamefont{E.}~\bibnamefont{Ott}}, \bibnamefont{and}
  \bibinfo{author}{\bibfnamefont{T.~M.} \bibnamefont{Antonsen}},
  \bibinfo{journal}{Chaos} \textbf{\bibinfo{volume}{21}},
  \bibinfo{pages}{023122} (\bibinfo{year}{2011}).

\bibitem[{\citenamefont{Hong and Strogatz}(2011{\natexlab{a}})}]{Hong:11a}
\bibinfo{author}{\bibfnamefont{H.}~\bibnamefont{Hong}} \bibnamefont{and}
  \bibinfo{author}{\bibfnamefont{S.~H.} \bibnamefont{Strogatz}},
  \bibinfo{journal}{Phys. Rev. Lett.} \textbf{\bibinfo{volume}{106}},
  \bibinfo{pages}{054102} (\bibinfo{year}{2011}{\natexlab{a}}).

\bibitem[{\citenamefont{Hong and Strogatz}(2011{\natexlab{b}})}]{Hong:11b}
\bibinfo{author}{\bibfnamefont{H.}~\bibnamefont{Hong}} \bibnamefont{and}
  \bibinfo{author}{\bibfnamefont{S.~H.} \bibnamefont{Strogatz}},
  \bibinfo{journal}{Phys. Rev. E} \textbf{\bibinfo{volume}{84}},
  \bibinfo{pages}{046202} (\bibinfo{year}{2011}{\natexlab{b}}).

\bibitem[{\citenamefont{Hong and Strogatz}(2012)}]{Hong:12}
\bibinfo{author}{\bibfnamefont{H.}~\bibnamefont{Hong}} \bibnamefont{and}
  \bibinfo{author}{\bibfnamefont{S.~H.} \bibnamefont{Strogatz}},
  \bibinfo{journal}{Phys. Rev. E} \textbf{\bibinfo{volume}{85}},
  \bibinfo{pages}{056210} (\bibinfo{year}{2012}).

\bibitem[{\citenamefont{Skardal et~al.}(2011)\citenamefont{Skardal, Ott, and
  Restrepo}}]{Skardal:11}
\bibinfo{author}{\bibfnamefont{P.~S.} \bibnamefont{Skardal}},
  \bibinfo{author}{\bibfnamefont{E.}~\bibnamefont{Ott}}, \bibnamefont{and}
  \bibinfo{author}{\bibfnamefont{J.~G.} \bibnamefont{Restrepo}},
  \bibinfo{journal}{Phys. Rev. E} \textbf{\bibinfo{volume}{84}},
  \bibinfo{pages}{036208} (\bibinfo{year}{2011}).

\bibitem[{\citenamefont{Lee et~al.}(2009)\citenamefont{Lee, Ott, and
  Antonsen}}]{Lee:09}
\bibinfo{author}{\bibfnamefont{W.~S.} \bibnamefont{Lee}},
  \bibinfo{author}{\bibfnamefont{E.}~\bibnamefont{Ott}}, \bibnamefont{and}
  \bibinfo{author}{\bibfnamefont{T.~M.} \bibnamefont{Antonsen}},
  \bibinfo{journal}{Phys. Rev. Lett.} \textbf{\bibinfo{volume}{103}},
  \bibinfo{pages}{044101} (\bibinfo{year}{2009}).

\bibitem[{\citenamefont{Sakaguchi and Kuramoto}(1986)}]{Sakaguchi:86}
\bibinfo{author}{\bibfnamefont{H.}~\bibnamefont{Sakaguchi}} \bibnamefont{and}
  \bibinfo{author}{\bibfnamefont{Y.}~\bibnamefont{Kuramoto}},
  \bibinfo{journal}{Prog. Theor. Phys} \textbf{\bibinfo{volume}{76}},
  \bibinfo{pages}{576} (\bibinfo{year}{1986}).

\bibitem[{\citenamefont{Daido}(1992)}]{Daido:92}
\bibinfo{author}{\bibfnamefont{H.}~\bibnamefont{Daido}},
  \bibinfo{journal}{Phys. Rev. Lett.} \textbf{\bibinfo{volume}{68}},
  \bibinfo{pages}{1073} (\bibinfo{year}{1992}).

\bibitem[{\citenamefont{Bonilla et~al.}(1993)\citenamefont{Bonilla,
  Perez~Vicente, and Rubi}}]{Bonilla:93}
\bibinfo{author}{\bibfnamefont{L.}~\bibnamefont{Bonilla}},
  \bibinfo{author}{\bibfnamefont{C.}~\bibnamefont{Perez~Vicente}},
  \bibnamefont{and} \bibinfo{author}{\bibfnamefont{J.}~\bibnamefont{Rubi}},
  \bibinfo{journal}{J. Stat. Phys.} \textbf{\bibinfo{volume}{70}},
  \bibinfo{pages}{921} (\bibinfo{year}{1993}).

\bibitem[{\citenamefont{Stiller and Radons}(1998)}]{Stiller:98}
\bibinfo{author}{\bibfnamefont{J.~C.} \bibnamefont{Stiller}} \bibnamefont{and}
  \bibinfo{author}{\bibfnamefont{G.}~\bibnamefont{Radons}},
  \bibinfo{journal}{Phys. Rev. E} \textbf{\bibinfo{volume}{58}},
  \bibinfo{pages}{1789} (\bibinfo{year}{1998}).

\bibitem[{\citenamefont{Daido}(2000)}]{Daido:00}
\bibinfo{author}{\bibfnamefont{H.}~\bibnamefont{Daido}},
  \bibinfo{journal}{Phys. Rev. E} \textbf{\bibinfo{volume}{61}},
  \bibinfo{pages}{2145} (\bibinfo{year}{2000}).

\bibitem[{\citenamefont{Stiller and Radons}(2000)}]{Stiller:00}
\bibinfo{author}{\bibfnamefont{J.~C.} \bibnamefont{Stiller}} \bibnamefont{and}
  \bibinfo{author}{\bibfnamefont{G.}~\bibnamefont{Radons}},
  \bibinfo{journal}{Phys. Rev. E} \textbf{\bibinfo{volume}{61}},
  \bibinfo{pages}{2148} (\bibinfo{year}{2000}).

\bibitem[{\citenamefont{Pikovsky and Rosenblum}(2008)}]{Pikovsky:08}
\bibinfo{author}{\bibfnamefont{A.}~\bibnamefont{Pikovsky}} \bibnamefont{and}
  \bibinfo{author}{\bibfnamefont{M.}~\bibnamefont{Rosenblum}},
  \bibinfo{journal}{Phys. Rev. Lett.} \textbf{\bibinfo{volume}{101}},
  \bibinfo{pages}{264103} (\bibinfo{year}{2008}).

\bibitem[{\citenamefont{Pikovsky and Rosenblum}(2011)}]{Pikovsky:11}
\bibinfo{author}{\bibfnamefont{A.}~\bibnamefont{Pikovsky}} \bibnamefont{and}
  \bibinfo{author}{\bibfnamefont{M.}~\bibnamefont{Rosenblum}},
  \bibinfo{journal}{Physica D} \textbf{\bibinfo{volume}{240}},
  \bibinfo{pages}{872} (\bibinfo{year}{2011}).

\bibitem[{\citenamefont{Ott and Antonsen}({2009})}]{Ott:09}
\bibinfo{author}{\bibfnamefont{E.}~\bibnamefont{Ott}} \bibnamefont{and}
  \bibinfo{author}{\bibfnamefont{T.~M.} \bibnamefont{Antonsen}},
  \bibinfo{journal}{{Chaos}} \textbf{\bibinfo{volume}{{19}}},
  \bibinfo{pages}{{023117}} (\bibinfo{year}{{2009}}).

\bibitem[{\citenamefont{Ott et~al.}({2011})\citenamefont{Ott, Hunt, and
  Antonsen}}]{Ott:11}
\bibinfo{author}{\bibfnamefont{E.}~\bibnamefont{Ott}},
  \bibinfo{author}{\bibfnamefont{B.~R.} \bibnamefont{Hunt}}, \bibnamefont{and}
  \bibinfo{author}{\bibfnamefont{T.~M.} \bibnamefont{Antonsen}},
  \bibinfo{journal}{{Chaos}} \textbf{\bibinfo{volume}{{21}}}
  (\bibinfo{year}{{2011}}).

\bibitem[{\citenamefont{Ott and Antonsen}({2008})}]{Ott:08}
\bibinfo{author}{\bibfnamefont{E.}~\bibnamefont{Ott}} \bibnamefont{and}
  \bibinfo{author}{\bibfnamefont{T.~M.} \bibnamefont{Antonsen}},
  \bibinfo{journal}{{Chaos}} \textbf{\bibinfo{volume}{{18}}},
  \bibinfo{pages}{{037113}} (\bibinfo{year}{{2008}}).

\bibitem[{\citenamefont{Petkoski et~al.}(2013)\citenamefont{Petkoski, Iatsenko,
  Basnarkov, and Stefanovska}}]{Petkoski:13}
\bibinfo{author}{\bibfnamefont{S.}~\bibnamefont{Petkoski}},
  \bibinfo{author}{\bibfnamefont{D.}~\bibnamefont{Iatsenko}},
  \bibinfo{author}{\bibfnamefont{L.}~\bibnamefont{Basnarkov}},
  \bibnamefont{and}
  \bibinfo{author}{\bibfnamefont{A.}~\bibnamefont{Stefanovska}},
  \bibinfo{journal}{Phys. Rev. E} \textbf{\bibinfo{volume}{87}},
  \bibinfo{pages}{032908} (\bibinfo{year}{2013}),
  \urlprefix\url{http://link.aps.org/doi/10.1103/PhysRevE.87.032908}.

\bibitem[{\citenamefont{Marvel et~al.}(2009)\citenamefont{Marvel, Mirollo, and
  Strogatz}}]{Marvel:09}
\bibinfo{author}{\bibfnamefont{S.~A.} \bibnamefont{Marvel}},
  \bibinfo{author}{\bibfnamefont{R.~E.} \bibnamefont{Mirollo}},
  \bibnamefont{and} \bibinfo{author}{\bibfnamefont{S.~H.}
  \bibnamefont{Strogatz}}, \bibinfo{journal}{Chaos}
  \textbf{\bibinfo{volume}{19}}, \bibinfo{pages}{043104}
  (\bibinfo{year}{2009}).

\bibitem[{\citenamefont{Strogatz and Mirollo}(1991)}]{Strogatz:91}
\bibinfo{author}{\bibfnamefont{S.~H.} \bibnamefont{Strogatz}} \bibnamefont{and}
  \bibinfo{author}{\bibfnamefont{R.~E.} \bibnamefont{Mirollo}},
  \bibinfo{journal}{J. Stat. Phys.} \textbf{\bibinfo{volume}{63}},
  \bibinfo{pages}{613} (\bibinfo{year}{1991}).

\bibitem[{\citenamefont{Omelchenko and Wolfrum}(2012)}]{Omelchenko:12}
\bibinfo{author}{\bibfnamefont{E.}~\bibnamefont{Omelchenko}} \bibnamefont{and}
  \bibinfo{author}{\bibfnamefont{M.}~\bibnamefont{Wolfrum}},
  \bibinfo{journal}{Phys. Rev. Lett.} \textbf{\bibinfo{volume}{109}},
  \bibinfo{pages}{164101} (\bibinfo{year}{2012}).

\bibitem[{\citenamefont{Omelchenko and Wolfrum}(2013)}]{Omelchenko:13}
\bibinfo{author}{\bibfnamefont{O.~E.} \bibnamefont{Omelchenko}}
  \bibnamefont{and} \bibinfo{author}{\bibfnamefont{M.}~\bibnamefont{Wolfrum}},
  \bibinfo{journal}{Physica D}  (\bibinfo{year}{2013}).

\bibitem[{sup()}]{suppmat}
\bibinfo{note}{For a dynamical version of the Figure 2, see
  \url{http://www.physics.lancs.ac.uk/research/nbmphysics/diats/km/}}.

\bibitem[{\citenamefont{Pikovsky and Rosenblum}(2009)}]{Pikovsky:09}
\bibinfo{author}{\bibfnamefont{A.}~\bibnamefont{Pikovsky}} \bibnamefont{and}
  \bibinfo{author}{\bibfnamefont{M.}~\bibnamefont{Rosenblum}},
  \bibinfo{journal}{Physica D} \textbf{\bibinfo{volume}{238}},
  \bibinfo{pages}{27} (\bibinfo{year}{2009}).

\bibitem[{\citenamefont{Filatrella et~al.}(2007)\citenamefont{Filatrella,
  Pedersen, and Wiesenfeld}}]{Filatrella:07}
\bibinfo{author}{\bibfnamefont{G.}~\bibnamefont{Filatrella}},
  \bibinfo{author}{\bibfnamefont{N.~F.} \bibnamefont{Pedersen}},
  \bibnamefont{and}
  \bibinfo{author}{\bibfnamefont{K.}~\bibnamefont{Wiesenfeld}},
  \bibinfo{journal}{Phys. Rev. E} \textbf{\bibinfo{volume}{75}},
  \bibinfo{pages}{017201} (\bibinfo{year}{2007}).

\bibitem[{\citenamefont{Rosenblum and Pikovsky}(2007)}]{Rosenblum:07}
\bibinfo{author}{\bibfnamefont{M.}~\bibnamefont{Rosenblum}} \bibnamefont{and}
  \bibinfo{author}{\bibfnamefont{A.}~\bibnamefont{Pikovsky}},
  \bibinfo{journal}{Phys. Rev. Lett.} \textbf{\bibinfo{volume}{98}},
  \bibinfo{pages}{064101} (\bibinfo{year}{2007}).

\bibitem[{\citenamefont{Baibolatov et~al.}(2010)\citenamefont{Baibolatov,
  Rosenblum, Zhanabaev, and Pikovsky}}]{Baibolatov:10}
\bibinfo{author}{\bibfnamefont{Y.}~\bibnamefont{Baibolatov}},
  \bibinfo{author}{\bibfnamefont{M.}~\bibnamefont{Rosenblum}},
  \bibinfo{author}{\bibfnamefont{Z.~Z.} \bibnamefont{Zhanabaev}},
  \bibnamefont{and} \bibinfo{author}{\bibfnamefont{A.}~\bibnamefont{Pikovsky}},
  \bibinfo{journal}{Phys. Rev. E} \textbf{\bibinfo{volume}{82}},
  \bibinfo{pages}{016212} (\bibinfo{year}{2010}).

\end{thebibliography}

\end{document}